\documentclass[11pt]{article}
\usepackage{epsfig}
\usepackage{varioref,exscale,latexsym,amsmath,amssymb}
\title{On the parameterization dependence of the energy momentum tensor and the metric}
\author{N.~E.~J Bjerrum-Bohr$^a$\\
John F. Donoghue$^b$, and Barry R. Holstein$^{b}$\\\\
$^a$ Department of Physics,\\
University of Wales Swansea, \\
Swansea, SA2 8PP, UK\\
and\\
Institute for Advanced Study,\\
Princeton, NJ 08540, USA\\\\
$^b$ Department of Physics\\
University of Massachusetts\\
Amherst, MA  01003, USA}
\begin{document}
\begin{titlepage}
\maketitle
\begin{abstract}
We use results by Kirilin to show that in general relativity the
nonleading terms in the energy-momentum tensor of a particle depends
on the parameterization of the gravitational field.  While the
classical metric that is calculated from this source, used to define
the leading long-distance corrections to the metric, also has a
parameteriztion dependence, it can be removed by a coordinate
change. Thus the classical observables are parameterization
independent. The quantum effects that emerge within the same
calculation of the metric also depend on the parameterization and a
full quantum calculation requires the inclusion of further diagrams.
However, within a given parameterization the quantum effects
calculated by us in a previous paper are well defined. Flaws of
Kirilin's proposed alternate metric definition are described and we
explain why the diagrams that we calculated are the appropriate
ones.
\end{abstract}

\end{titlepage}

In \cite{bbdh1}, we calculated the long distance one loop
corrections to the energy momentum tensor of spin zero and spin
one-half particles due to graviton loops and used this result to
derive the long distance corrections to the metric. The diagrams are
shown in Figure 1. This procedure reproduces the leading classical
nonlinearities in the Schwarzschild and Kerr metrics - in harmonic
gauge - and produces novel quantum corrections linear in $\hbar$.
\begin{figure}[h]
\begin{center}
\epsfig{file=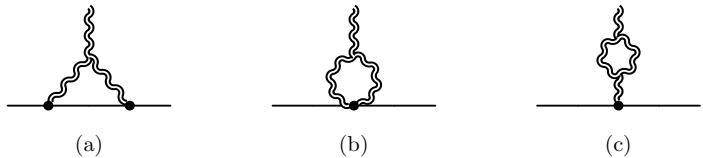,height=3cm,width=10cm}
\caption{Feynman diagrams used to calculate the long-distance
corrections to the energy-momentum tensor and the metric of a scalar
particle. Here
 the doubly wiggly lines represent gravitons.}
\end{center}
\end{figure}

As noted in our paper, this calculation is not a full quantum
calculation. The metric is not a fully quantum concept and further
diagrams are needed in order to produce a proper quantum amplitude.
Indeed, in the companion paper \cite{bbdh2}, we completed the
calculation of the classical and quantum corrections to the
gravitational scattering amplitude. The full set of diagrams is
shown in Figure 2. This result has also been calculated by
Khriplovich and Kirilin \cite{kk} and our results agree.

\begin{figure}[h]
\begin{center}
\epsfig{file=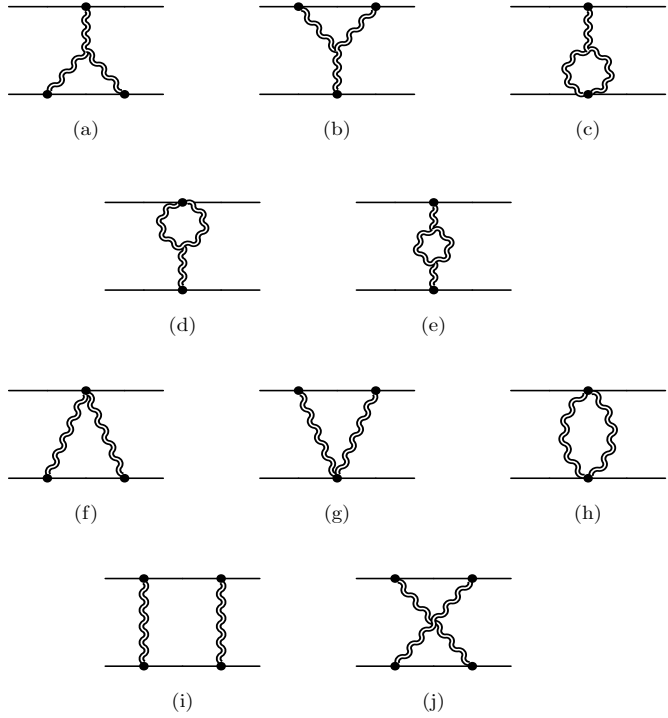,height=10cm,width=10cm}
\caption{The diagrams used to calculate the long-distance
corrections to the scattering amplitude.} 
\end{center}
\end{figure}
In the preceding comment \cite{wrong}, Kirilin criticizes our choice
of diagrams to include in the definition of the metric, and proposes
a metric derived from the work on the scattering amplitude
\cite{kk}. This criticism is based on a calculation which shows that
the metric derived from Fig 1 is not invariant under the
reparameterization of the graviton field. This is interesting and
appears to be correct - we will comment further below.  It implies
that when discussing the quantum corrections to the metric one must
specify not only the gauge but also the field parameterization.
However, we will argue that our definition of the metric is still to
be preferred and that the one proposed by Kirilin has a number of
flaws.

Kirilin considers the family of parameterizations of the metric
field
\begin{equation}
g_{\mu\nu}= \eta_{\mu\nu} + h_{\mu\nu}-\frac{a}{4}
h_{\mu\lambda}h^\lambda_{\nu}  ~~~ ,
\end{equation}
where $a$ is a free parameter. Changes in $a$ lead to changes in the
various vertices entering the Feynman diagrams \cite{wrong}, in
particular the triple graviton coupling and also the coupling of two
gravitions to the matter field. For example, when the results of
\cite{wrong} are applied to the energy momentum tensor of a scalar
particle
\begin{equation}
<p_2|T_{\mu\nu}(x)|p_1>={e^{i(p_2-p_1)\cdot x}\over
\sqrt{4E_2E_1}}\left[2P_\mu P_\nu F_1(q^2)+(q_\mu
q_\nu-\eta_{\mu\nu} q^2)F_2(q^2)\right] ~~~,
\end{equation}
with $P_\mu =(p_1+p_2)_\mu/2$ and $q_\mu=(p_1-p_2)_\mu$, the
consideration of the diagrams of Fig 1, a,b, show that the matrix
element of $T_{\mu\nu}$ depends on the parameterization of the
gravitational field. Specifically
\begin{eqnarray}
F_1(q^2)&=&1+{Gq^2\over \pi}(-{3\over 4}\log{-q^2\over m^2}+
{1\over 16}{\pi^2 m\over \sqrt{-q^2}})+\ldots\nonumber\\
F_2(q^2)&=&-{1\over 2}+{Gm^2\over \pi}((-2-(\frac{11}{3}a
+a^2))\log{-q^2\over m^2}+{7+4a \over 8}{\pi^2m\over
\sqrt{-q^2}})+\ldots  ~~~ .
\end{eqnarray}
Here we have displayed only the nonanalytic terms that give long
range modifications.  As shown in \cite{bbdh1}, the square-root
terms lead to classical corrections, while the logarithms lead to
quantum corrections. We notice that the form factor $F_1$ is
unaffected, while $F_2$ is modified in {\it both} the classical and
quantum components.

It is interesting that the energy-momentum tensor depends on the
gravitational field parameterization. In non-gravitational theories,
the energy and momentum are well defined quantities and such a
matrix element suffers no ambiguities. However, in general
relativity, Noether's theorem does not produce a well defined energy
because the action is invariant under general coordinate
transformations. This feature is manifest in the one loop
corrections to the particle's energy, and also in the shift of the
gravitational vertices. For example, the $a$-dependent modification
of the classical component arises from the energy and momentum
contained in the classical gravitational field surrounding a
particle, which in the loop expansion is described by Fig 1a
\cite{classical}. However, the energy-momentum tensor for the
gravitational field is a pseudo-tensor\footnote{It is equivalent to
the triple graviton coupling when one of the gravitons is taken as
an external gravitational field.} that depends on the field
parameterization (i.e. on $a$). The variation in that tensor means
that the amount of energy and momentum that is carried in the
classical field also varies with the parameterization. The variation
in the quantum component comes from this effect, plus an additional
change which results from a shift in the two graviton ($\phi\phi
hh$) coupling, which is involved in Figs. 1b.

Using the energy momentum tensor as the source, we calculated in
\cite{bbdh1} the change in the metric by inverting the linear
gravitational field equations in harmonic gauge,
\begin{eqnarray}
h_{\mu\nu}(x) &=& -16\pi G \int d^3y D(x-y) (T_{\mu\nu}(y)-{1\over
2}\eta_{\mu\nu}
   T(y))  \nonumber\\
&=& -16\pi G \int {d^3q \over (2\pi)^3}
e^{i\vec{q}\cdot\vec{r}}{1\over \vec{q}^2}(T_{\mu\nu}(q)-{1\over
2}\eta_{\mu\nu}
   T(q)) ~~~ ,
\end{eqnarray}
and this result then also depends on the parameterization\footnote{
For reasons described in our paper, we also included the vacuum
polarization diagram in the metric, but this only influences the
quantum corrections and does not remove the dependence on the
parameterization.}.

Let us here comment particularly on the classical component, for
which this procedure is very well defined. The result for the
classical fields $h_{\mu\nu}$ are found to be
\begin{eqnarray}
h_{00} &=& -\frac{2Gm}{r}\left[1
-(1+\frac{a}{2})\frac{Gm}{r}+...\right]
\nonumber\\
h_{ij} &=& -\frac{2Gm}{r}\left[\delta_{ij} +\frac{Gm}{2r}\left(
(\delta_{ij} +\frac{r_ir_j}{r^2})- a(3\delta_{ij}
-4\frac{r_ir_j}{r^2})\right)+...\right]
\end{eqnarray}
Since classical observables cannot depend on the field
parameterization, it is interesting to see how the parameterization
dependence disappears. Let us form the total metric from
$h_{\mu\nu}$ through the use of Eq 1. We find
\begin{eqnarray}
g_{00} &=& 1 -\frac{2Gm}{r}\left[1 -\frac{Gm}{r}+...\right]
\nonumber\\
g_{ij} &=& -\delta_{ij} -\frac{2Gm}{r}\left[\delta_{ij}
+\frac{Gm}{2r}\left( (\delta_{ij} +\frac{r_ir_j}{r^2})- 2
a(\delta_{ij} -2\frac{r_ir_j}{r^2})\right)+...\right]
\end{eqnarray}
As noted in \cite{wrong}, the iterated paramterization dependence
from Eq 1 cancels the other $a$ dependence in $g_{00}$, leaving a
parameterization independent result. However, the same does not
occur in $g_{ij}$. This correction is interesting. In the first
place, it shifts the metric away from the harmonic gauge. The
correct form of the harmonic gauge metric, i.e. satisfying
$g^{\mu\nu}\Gamma^\lambda_{\mu\nu} =0$, is recovered in the $a=0$
limit. For a general parameterization we find
\begin{equation}
g^{\mu\nu}\Gamma^i_{\mu\nu} = -2a \frac{G^2m^2}{r^4}r^i
\end{equation}
So we see that the parameterization change is also associated with a
change of gauge. However, because it is simply a gauge change, a
change of coordinates of the form
\begin{equation}
x^i = (1 -a\frac{G^2m^2}{2r^2})x^i
\end{equation}
can be used to bring convert the metric back to harmonic gauge form.
Thus classical observables are invariant under the parameterization
change.

The quantum parameterization dependence cannot be removed by a
change in coordinates. The consequence of this residual dependence
is that, when performing the calculation of a full quantum
amplitude, one must take care to use the same field paramaterization
for the entire calculation. This aspect is not unique to gravity.
For example, in the internucleon potential, a reparameterization of
the pion field modifies the central nuclear potential \cite{friar},
although the sum of pionic exchanges and the central potential is
invariant. In both cases, a consistent calculation of intermediate
results must involve a full specification of the field variables,
which in the gravitational case means both the gauge and the field
definition.

Kirilin \cite{wrong} proposes instead to use a different definition
of the metric, based on the full set of diagrams shown in Fig 2. We
do not feel that this alternative definition is acceptable. Flaws of
this set of diagrams include:

1) These diagrams do not reproduce the classical corrections to the
metric. Instead they reproduce the classical terms in the
post-Newtonian potential. These are {\it not} the same. As discussed
above, our diagrams capture the correct classical metric
corrections. There are additional classical corrections are found in
diagrams 2.f-j, shifting the result to the post-Newtonian potential
rather than the metric. Kirilin's set of diagrams does not give the
correct classical metric.

2) The diagrams of Fig. 2 form a gauge-invariant set. In contrast,
the metric depends very explicitly on the gauge for the
gravitational field. In \cite{kk}, the authors correct for this
deficiency by hand, but the correct gauge dependence is not a
feature of either the classical or quantum components of these
diagrams.

3) The diagrams proposed by Kirilin depend on the existence of a
second non-relativistic particle in the diagrams. For example, it
includes the vertex correction to the gravitational vertex of the
{\it other} particle, Fig 2b. The classical corrections depend on
the mass of the other particle, and there exist terms in the
interaction that depend on the spin of the other particle. There is
also no indication that the same result would be obtained if the
other particle were an ultra-relativistic particle such as a photon.
Of course, one can deal with these features in an ad-hoc way by
taking one mass much larger than the other and excising the spin
dependence by hand. However, the basic feature of a metric is that
it is the property of a single object independent of the existence
of a second body. These diagrams do not have that property.

4) The mechanism that is required in order to obtain the invariance
of the classical observables, described above and in \cite{wrong},
does not work with the classical part of the diagrams of Fig. 2 but
does if one uses those of Fig. 1.

5) In the case of electromagnetic corrections to the metric, there
is no ambiguity due to reparameterization of the gravitational
field, because gravity in this case is purely classical. Here, both
the diagrams corresponding to Fig 1 and Fig 2 have been calculated
\cite{photonmetric, photonscat}\footnote{The vacuum polarization
diagram is not relevant in this case.}, and it has been demonstrated
that the diagrams of Figure 1 give the correct metric. Kirilin's
procedure in this case would be incorrect. The above comments 1-4
also apply to the photonic calculation.

Our definition of the metric includes the full set of one-loop
diagrams for the gravitational field around a single body. It
reproduces correctly the classical terms in the metric, and displays
the correct gauge dependence. It parallels the well-defined
corrections due to photonic loops. Because of the parameterization
dependence of the energy-momentum matrix element, the metric also
displays a parameterization dependence. However, the classical
dependence within these diagrams is exactly what is required in
order that classical observables be parameterization independent.
Kirilin's alternate definition is shown to be inappropriate for the
classical components. While the metric is not a full quantum
calculation, within a given parameterization our definition can be
completed to give the calculation of the full quantum amplitude. In
summary, we stand by our choice of diagrams describing the classical
and quantum corrections to the metric.

\section*{Acknowledgements}

The research of JFD and BRH has been supported in part by the U.S.
National Science Foundation under grant NSF 05-53304 and that of
NEJBB by the PPARC of the UK and the Department of Energy under
grant DE-FG02-90ER40542.

\end{document}